# Amplitude modulated, by M1, Earth's oscillating (T = 1 day) electric field triggered by K1 tidal waves. Its relation to the occurrence time of large EQs.

Thanassoulas[1], C., Klentos[2], V., Verveniotis, G[3].

1. Retired from the Institute for Geology and Mineral Exploration (IGME), Geophysical Department, Athens, Greece.
   e-mail: thandin@otenet.gr - URL: www.earthquakeprediction.gr

2. Athens Water Supply & Sewerage Company (EYDAP),
   e-mail: klenvas@mycosmos.gr - URL: www.earthquakeprediction.gr

3. Ass. Director, Physics Teacher at 2nd Senior High School of Pyrgos, Greece.
   e-mail: gver36@otenet.gr - URL: www.earthquakeprediction.gr

**Abstract.**

Starting from the observation that quite often the Earth's oscillating electric field varies in amplitude, a mechanism is postulated that accounts for these observations. That mechanism is the piezoelectric one driven by the **M1** and **K1** tidal components. It is demonstrated how the system: piezoelectricity triggered in the lithosphere – **M1** and **K1** tidal components is activated and produces the amplitude modulated Earth's oscillating electric field. This procedure is linked to the strain load conditions met in the seismogenic area before the occurrence of a large EQ. Peaks of the oscillating Earth's electric field are tightly connected to the **M1** peak tidal component and to the timing of the occurrence of large EQs. Typical examples from real recordings of the Earth's oscillating electric field, recorded by the **ATH (Greece)** monitoring site, are given in order to verify the postulated detailed piezoelectric mechanism.

## 1. Introduction.

Generally, an earthquake occurs when the accumulated strain of the rock formation of the seismogenic area exceeds its fracturing level. What is very interesting is what happens in the seismogenic area during its critical state just before the occurrence of the seismic event. During this stage of strain load, only a small amount of extra strain is required to overcome the fracturing threshold of the rock formation in order to trigger the future earthquake. A mechanism which can provide such an extra strain load is the tidal waves generated by the Sun and Moon interaction upon the Earth's surface.

The Earth-tides mechanism was early recognized as a potential trigger for the occurrence of strong earthquakes. This approach was followed in studying the time of occurrence of EQs and their correlation to Earth-tides. Knopoff (1964), Shlien (1972), Heaton (1982) and Shirley (1988) suggested the Earth-tides as a triggering mechanism of strong EQs, Yamazaki (1965, 1967), Rikitake et al. (1967) studied the oscillatory behaviour of strained rocks, due to Earth-tides, while Ryabl et al. (1968), Mohler (1980) and Sounau et al. (1982) correlated Earth-tides to local micro-earthquakes and aftershock sequences.

A detailed account of the holding mechanism that generates the required extra strain load in order to trigger the future earthquake has been presented by Thanassoulas (2007). The strain load of the seismogenic area, at any time, consists of a linear component due to very long term (in terms of years) strain increase and the superimposed tidal oscillating ones. Since the tidally induced strain variation follows the tidal waves and because the tidal periods involved in this mechanism are much more shorter than the corresponding to the long term one (if we consider it as an oscillatory strain increase of period of some decades of years – being the repetition rate of large EQs in Greece) then it is clear that the fracturing threshold level of the rock formation in the seismogenic area will be reached theoretically at the peaks of the tidal components. Therefore, large EQs are expected to occur at the peaks of the tidal waves.

Two statistical tests were run for the validation of afore mentioned mechanism (Thanassoulas, 2007). The first test compared the timing of the tidal wave peaks of the **M1** component (**T = 14 days**, moon declination) to the time of occurrence of large EQs (**Ms >= 5.5R**) for the time period from 1995 to 2001. The number of seismic events that took place during this period of time is 40. The "by chance" for an EQ to occur in any day between two successive tidal peaks (7 days) is **14.28%**. This test resulted in **39.47%** coincidence of the studied EQs with the exact tidal peak. If a window of +/- 1 day deviation from the tidal peak was tolerated then the coincidence raised to **50%**. The second test compared the timing of the tidal wave peaks of the **K1 (T = 24 hours, lunisolar)** to the occurrence time of EQs of **Ms >= 6.0R** for the time period from 1964 to 2001. The number of seismic events that took place during this period of time is 70. The "by chance" for an EQ to occur in any time between two successive tidal peaks (12 hours) is **16.1%**. The test results have shown that by accepting an occurrence window that deviates at 1 hour only from the tidal peak then a percentage of **37.1%** of the studied EQs fall in this window. If the magnitude of the studied EQs is raised to **Ms >= 6.5R** then the 50% of the studied EQs fall within the time window. From both tests run it is evident the strong control of the tidal waves upon the triggering time of large EQs.

Besides the strong statistical relation of the occurrence time of large EQs with the tidal waves, the latter present a very important feature. The tidal waves cause the generation of oscillating seismic precursory signals in the lithosphere. Such signals of 1 day's (**T = 24 hours**) period were reported by Thanassoulas (1982) and Thanassoulas et al. (1986, 1993).

Examples of preseismic oscillating signals of larger period (**T = 14 days**) have been presented by Thanassoulas (2007). Furthermore, the tidal waves give rise to the generation of the Seismic Electric Signals (**SES**) (Thanassoulas, 2008). The latter signals are closely clustered to the tidal wave peak of 14 days period (**M1**).

It seems that the large / small scale piezoelectric model (Thanassoulas, 2008a), adopted for the generation of preseismic electric signals in the lithosphere, plays a very important role in understanding the observed electric side effects prior to the seismogenesis.

Consequently, it is very interesting to analyze how the oscillating Earth's electric field relates to the **T = 14 days** tidal waves (**M1**) and how the time of occurrence of large EQs is connected to the same tidal waves too. To this end, the corresponding tidal wave peaks will be compared to the Earth's oscillating electric field recorded by the **ATH** (Athens) monitoring site ([www.earthquakeprediction.gr](www.earthquakeprediction.gr)) for a time period of 27 months from 20080101 to 20100410 (YYYYMMDD mode) and to the large seismicity that took place at the same period of time.

## 2. Data presentation and analysis.

Locally horizontal extension and compression takes place (Thanassoulas 2007) at a seismogenic area in various frequency modes as long as the lithosphere oscillates, triggered by the corresponding components of the tidal waves. The form of these lithospheric oscillations is the same as the triggering mechanism mode. In this case it is the Earth's gravity field oscillation due to the Moon – Sun – Earth, combined at any moment, attraction. Therefore, it is possible to figure out the lithospheric oscillations by analyzing the gravity field oscillating component at any place. The latter has been used extensively for "tidal corrections" applied on the gravity data collected during at any gravity survey.

### 2.1 Calculating the M1 tidal wave component.

The gravity field oscillation is calculated, at first, at 1 minute sampling interval, by using the Rudman et al. (1977) method. A sample of a three days period is presented in the following figure (1). The Rudman's method calculates the **vertical component** of the oscillating gravity field since it was originally written for gravity survey applications (tidal corrections).

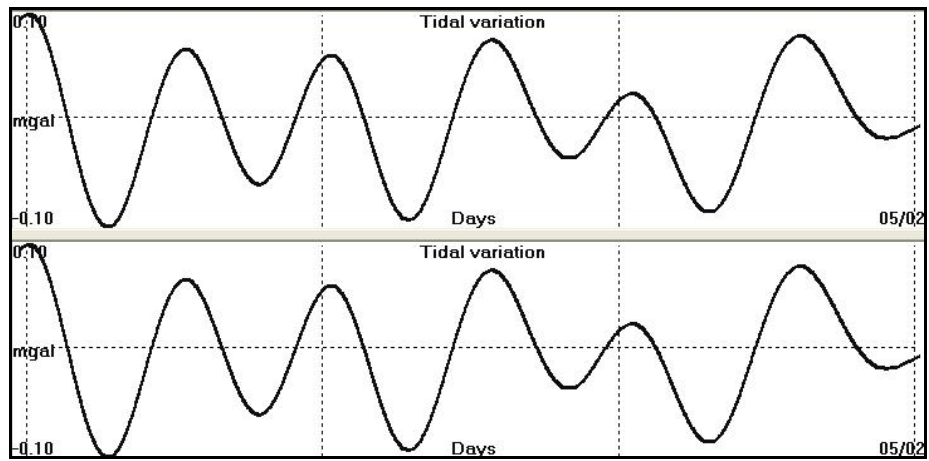

Fig. 1. Graph represents the daily tidal variation (vertical component in mgals) for three consecutive days.

The most evident tidal components are the ones of the **T = 12hours** range (**K2, N2, M2, S2**) while the **T = 24** hours range components (**K1, P1**) are shown too. A longer tidal sample is given in the following figure (2).

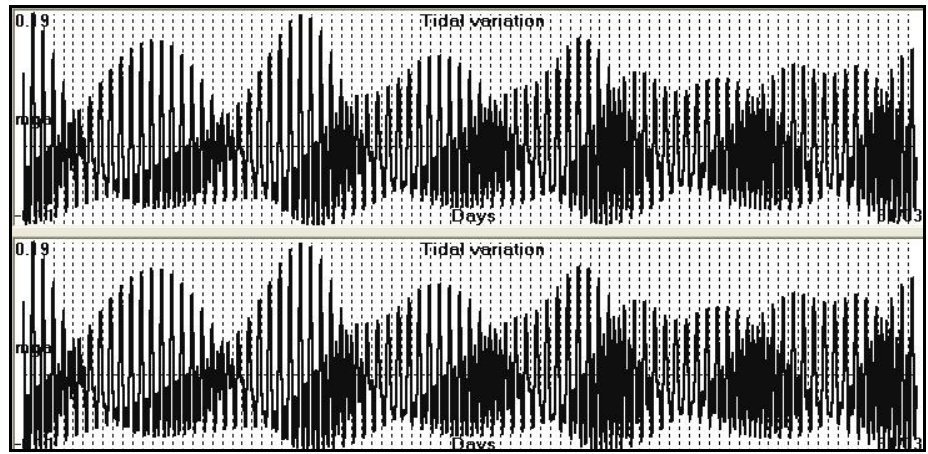

Fig. 2. Daily tidal variation determined for the period from 20100101 – 20100331

Although the graph is quite dense, due to the 1 minute sampling interval, it is evident that components of much larger periods than that of one day's do exist. Therefore, aiming into isolating these large period components, a low-pass filtering procedure (Bath 1974; Kulhanec 1976) is applied upon the tidal data. The latter is achieved by re-sampling the data at a sampling rate of the lowest unwanted period (Nyquist upper band limit). In this case the tidal data were re-sampled at a day's sampling interval. The results of this filtering operation are shown in the following figure (3).

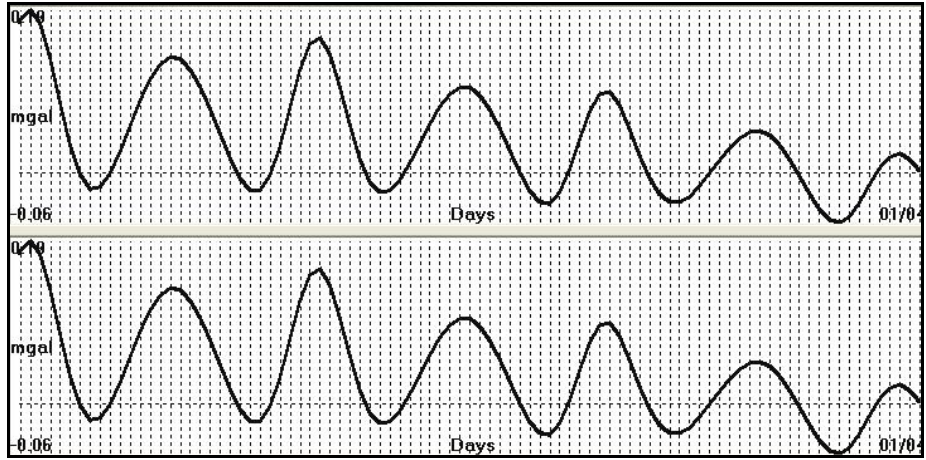

**Fig. 3. Re-sampling of the daily tidal data (20100101 – 20100331) of figure (2) at 1 sample per day. Frequency components of T < 1 day are totally rejected while components of T = 14 days are clearly shown.**

**2.2. Study data set (Time span: 20080101 – 20100410).**

The study period spans from 20080101 to 20100410 (YYYYMMDD mode). The tidal gravity variation, calculated at 1 minute sample interval by the Rudman's method and for the entire study period, was re-sampled at a day's sampling interval. The results are presented in the following figure (4).

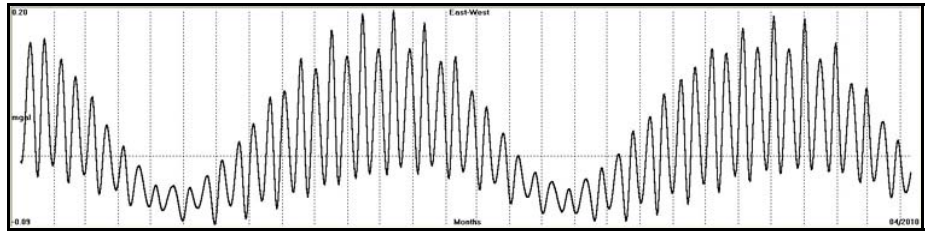

**Fig. 4. Re-sampled Tidal variation calculated for 20080101 – 20100410 period of time.**

Figure (4) visualizes quite well a: the yearly tidal variation of the gravity field and b: the M1 tidal component (T=14 days). Next, the power spectrum of the data of figure (4) has been calculated, by Fast Fourier Transformation (FFT), in order to investigate, in more details, the frequency content of the data of figure (4). The power spectrum is presented in the following figure (5).

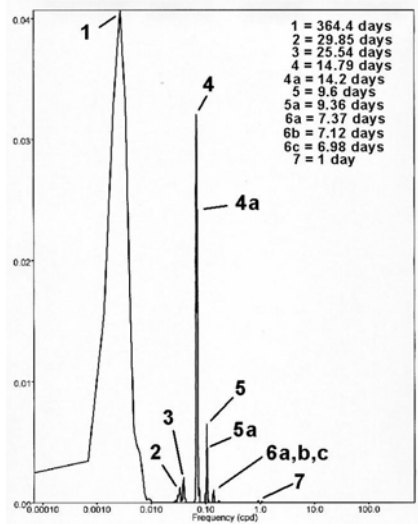

**Fig. 5. Power spectrum of the data of figure (4) is presented. The upper right table indicates the various identified frequency (periods) components. 4 and 4a correspond to M1 (moon declination) and 3 corresponds to O1 (lunar declination).**

What is interesting in figure (5) is the large amplitudes tidal gravity components of the no. 1 (yearly) and of 4 and 4a (M1 moon declination) component with period in the range of 14 days. The 24 hours period component (no. 7) has completely diminished.

A simpler approach for determining the M1 component is to calculate it directly from figure (4) by measuring the elapsed time (half period) between two consecutive tidal peaks. The application of this procedure upon the data of figure (4) resulted in the following graph of figure (6).

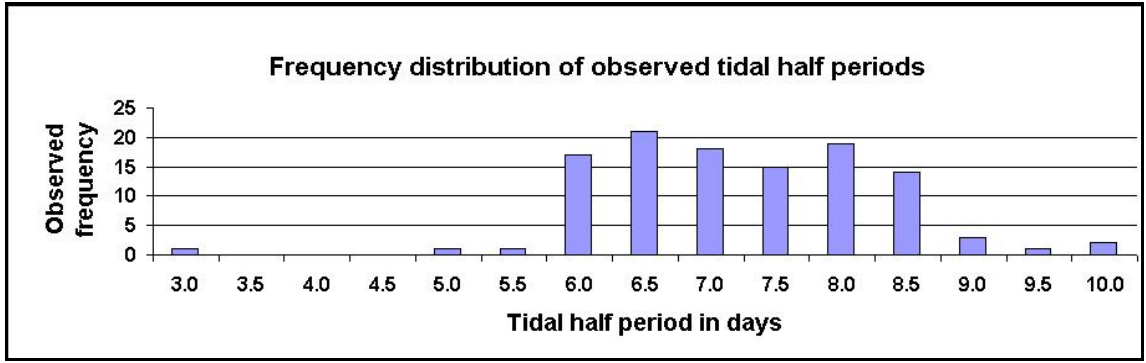

Fig. 6. Half period of the 14 days tidal oscillation determined from actual tidal data of figure (4).

The calculated half periods are clustered (in their vast majority) between 6 and 8.5 days. Consequently, the resulted average value for M1 is: T = 14.4778 days. The latter complies quite well with the results of the FFT analysis which provided: T = 14.79 days and T = 14.2 days for M1 (see fig. 5). The observed dispersion of the calculated half period values could be attributed to the simultaneous presence of slightly different than the M1 periods in the used data or errors introduced by the filtering (re-sampling) used procedure.

A more sophisticated explanation could be as follows: The M1 and K1 tidal components interact in the lithospheric system, which oscillates in a non-linear mode due to its non-linear mechanical deformation. Therefore, through the non-linearity of the lithospheric system, amplitude modulation is activated. In this case the "carrier" (in terms of amplitude modulated wave broadcasting terminology) is represented by the M1 tidal component while the modulating wave (modulating audio frequencies) are represented by the K1 tidal component. The results of this operation are very well known to physicists. A frequency spectrum is generated that consists of the "carrier" M1 and the generated sidebands: M1 + K1, and M1 - K1 (Orr, 1967). Consequently, for the case of the interaction of the M1 and K1 tidal components, it is expected that tidal waves (of small amplitude) will be generated that will cover a wider range higher and lower (upper and lower sidebands) from the central "carrier" tidal component M1. The latter mechanism complies quite well with the results already presented in figure (6).

## 2.3. Postulated piezoelectric model.

The piezoelectric model that was postulated by Thanassoulas et al. (1986, 1993) justified the generation of oscillating (T = 1 day) electric signals as long as the seismogenic area had entered its last deformation phase and particularly the non-linear deformation part of the stress – strain curve. By that time it was initially assumed that the stress / strain load was increasing slowly due to plate tectonic motion. After the study of a long (some years) recording of the Earth's electric field it was observed that there were periods with no oscillation amplitude increase, periods of amplitude oscillation increase and periods of amplitude oscillation increase that was followed by the occurrence of a large earthquake. Therefore, the piezoelectric model should be studied and analyzed in further details in order to account for all these type of observed electric field oscillations.

In particular, the stress – strain increase of the seismogenic area was studied in terms of the M1 tidal component. Three phases can be distinguished in this mechanism a: phase – 1, b: phase – 2 and c: phase – 3.

### 2.3.1. Piezoelectric phase – 1.

In phase – 1, the stress – strain load (A) of the seismogenic area is far from triggering the activation of the piezoelectric mechanism. Although the stress load increases linearly due to the plate motion, it simultaneously swings around (A) due to the M1 tidal component but it never reaches the critical stress load level (1) so that it triggers the activation of the piezoelectric mechanism. Phase – 1 is presented in figure (7). Consequently, the Earth's oscillating electric field will present rather constant amplitude due to existent random ambient electric noise. A sample of such a recording is presented in the following figure (8).

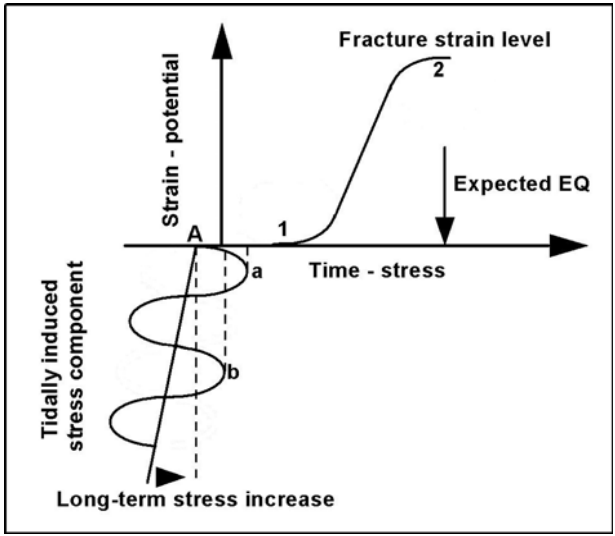

Fig. 7. Seismogenic area is in **phase – 1** stress – strain load conditions. The temporary stress load **(A)** increases linearly due to the long-term stress increase while at the same time oscillates due to the **M1** tidal component. In this case no piezoelectric potential is activated since the total (long term increase plus oscillating) at its maxima **(a, b)** are well before the critical point **(1)** when piezoelectricity is activated. The **K1** tidal component cannot trigger daily Earth's electric field oscillations.

**A typical recording of such a case is presented in figure (8).**

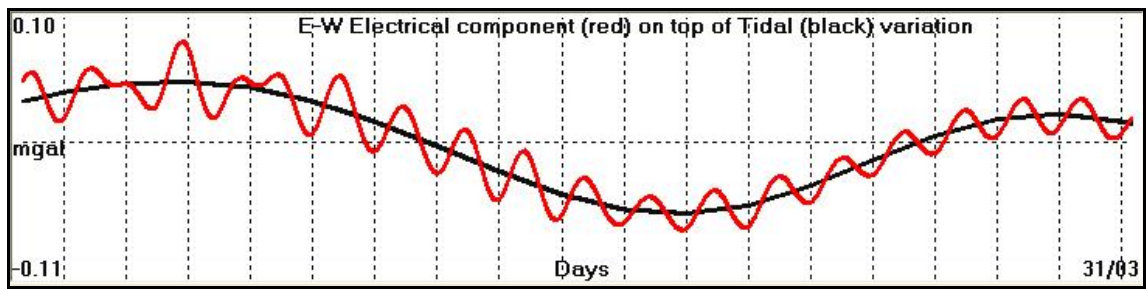

Fig. 8. Recording of the Earth's electric field for the period 20100313 – 20100331 (YYYYMMDD mode). The black line represents the **M1** tidal wave while the red line represents the Earth's oscillating electric field. No amplitude increase of the oscillating electric field is observed during the **M1** tidal peak.

### 2.3.2. Piezoelectric phase – 2.

In **phase – 2**, the stress – strain load **(B)** is just before activating the piezoelectric mechanism. Consequently, as long as the total stress – strain of the seismogenic area swings around **(B)**, it partially activates the piezoelectric mechanism. In such a case, daily oscillations of the Earth's electric field are observed, due to **K1**, but with initially increasing and afterwards decreasing amplitude. The increase is observed as long as the total stress – strain **(B)** enters the stress – strain piezoelectric curve, while it decreases and probably diminishes in the opposite direction. This phenomenon is totally controlled by the **M1** tidal component. **Phase – 2** is presented in figure (9).

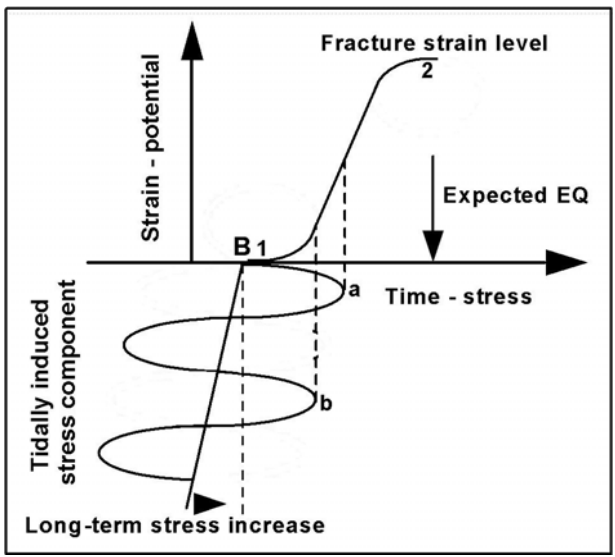

Fig. 9. Seismogenic area is in **phase – 2** stress – strain load conditions. The temporary stress load **(B)** increases linearly due to the long-term stress increase while at the same time oscillates due to the **M1** tidal component. In this case the piezoelectric potential is activated since the total (long term increase plus oscillating) at its maxima **(a, b)** are well beyond the critical point **(1)** when piezoelectricity is activated and **K1** tidal component triggers daily oscillations of the Earth's electric field.

A typical recording of such a case is presented in figure (10).

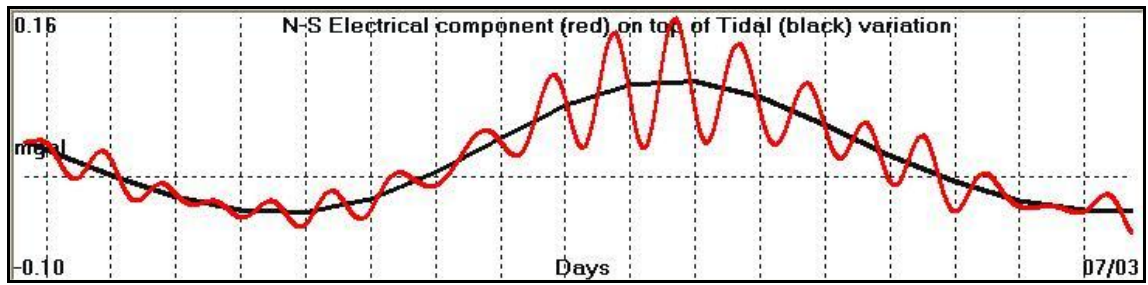

**Fig. 10.** Recording of the Earth's electric field for the period 20100218 - 20100310 (YYYYMMDD mode). The black line represents the M1 tidal wave while the red line represents the Earth's oscillating electric field due to K1. A well observed amplitude increase of the oscillating electric field is observed during the M1 tidal peak.

It must be commented here that although an increasing in amplitude oscillation is observed, since the strain load does not reaches the fracturing level of the seismogenic rock formation then no earthquake is triggered.

### 2.3.3. Piezoelectric phase – 3.

In **phase – 3**, the stress – strain load **(C)** is in the region **(1, 2)** when the piezoelectric mechanism has been activated. Consequently, the oscillating stress – strain load generates oscillating electric signals of drastically increasing amplitude. In such a case it is very probable that some peak (within a couple of days) of the oscillating stress – strain load will exceed the fracturing level of the rock formation of the seismogenic area and therefore an earthquake will take place. **Phase – 3** is presented in the following figure (11).

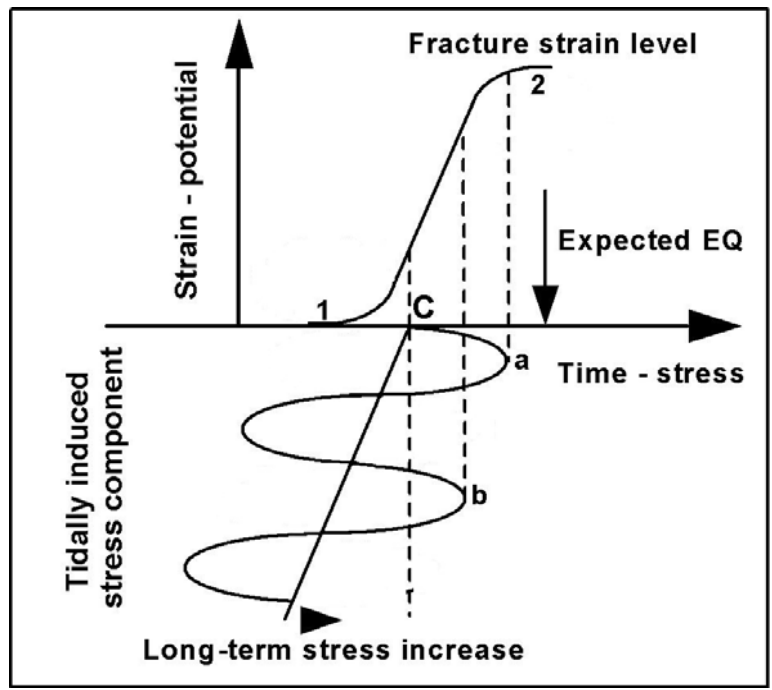

**Fig. 11.** Seismogenic area is in **phase – 3** stress – strain load conditions. The temporary stress load **(C)** increases linearly due to the long-term stress increase, while at the same time oscillates due to the M1 tidal component. In this case the piezoelectric potential is activated since the total (long term increase plus oscillating) at its maxima **(a, b)** are well beyond the critical point **(1)** when piezoelectricity is activated. Simultaneously, daily oscillations of the Earth's electric field are triggered due to K1. Consequently it is highly probable that the fracturing strain level of the rock formation of the seismogenic area will be reached (in a couple of days) and therefore an earthquake will take place.

A typical recording of such a case is presented in figure (12).

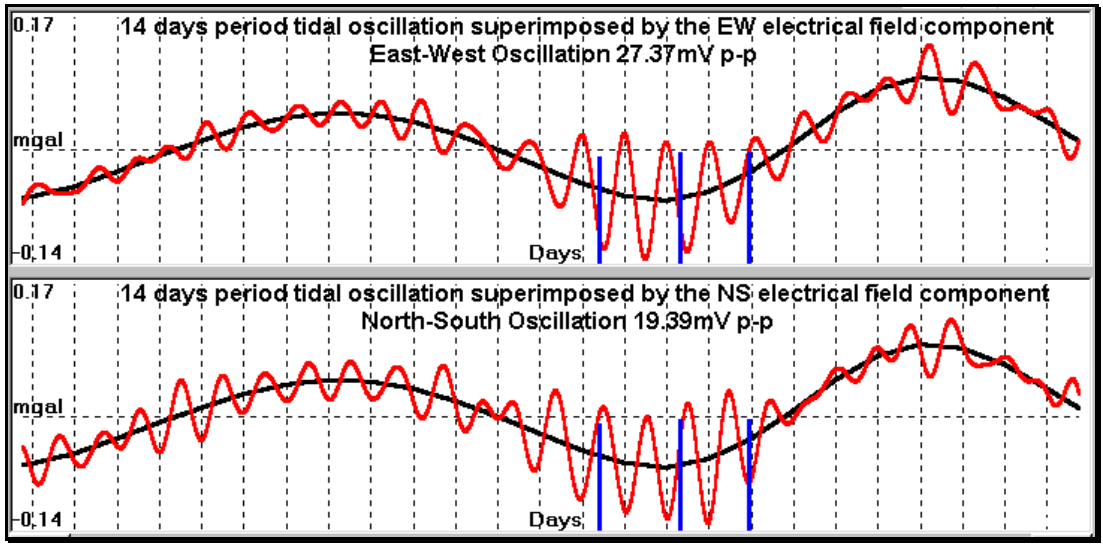

**Fig. 12. Recording of the Earth's electric field for the period 20020811 – 20020904 (YYYYMMDD mode). The black line represents the M1 tidal wave while the red line represents the Earth's oscillating electric field. Apart from the observed increase of the amplitude of the oscillating Earth's electric field due to K1, during the peak of the M1 tidal wave, three earthquakes of M = 5.1R (Thanassoulas et al. 2003) took place during the peak tidal – electric field period. The blue bars indicate the time of occurrence of these earthquakes.**

It must be remembered that the **T = 24hours** oscillation of the Earth's electric field, observed during the stress loading of the seismogenic area, is generated due to the **K1** tidal wave oscillating component which swings in a daily mode the stress load **(B, C)** simultaneously with the **M1** (see figures 9, 11 and Thanassoulas et al. 1986, 1993). In conclusion, the daily Earth's oscillating electric field is considered as the "carrier" being generated by the **K1** tidal component, as long as it occurs during the **phase – 3** of the piezoelectric mechanism, which carrier in turn is "amplitude modulated" by the **M1** tidal component. Moreover, the strain level of the fracturing of the rock formation in the majority of the cases occurs when both **(M1, K1)** tidal peaks are in phase and therefore it is possible to predetermine the timing within a day of the occurrence of a large EQ (Thanassoulas et al. 2010).

One more typical example of the presented mechanism that generates oscillating signals of **T = 1 day** due to M1 and K1 is shown in the following figure (13). The recording spans from 20100220 to 20100324. The upper graph represents the **E – W** electric field component while the lower one represents the **N – S** electric component. Two successive Earth's oscillating electric field peaks are clearly observed (on **N-S**) and coincide to the **M1** tidal components peaks.

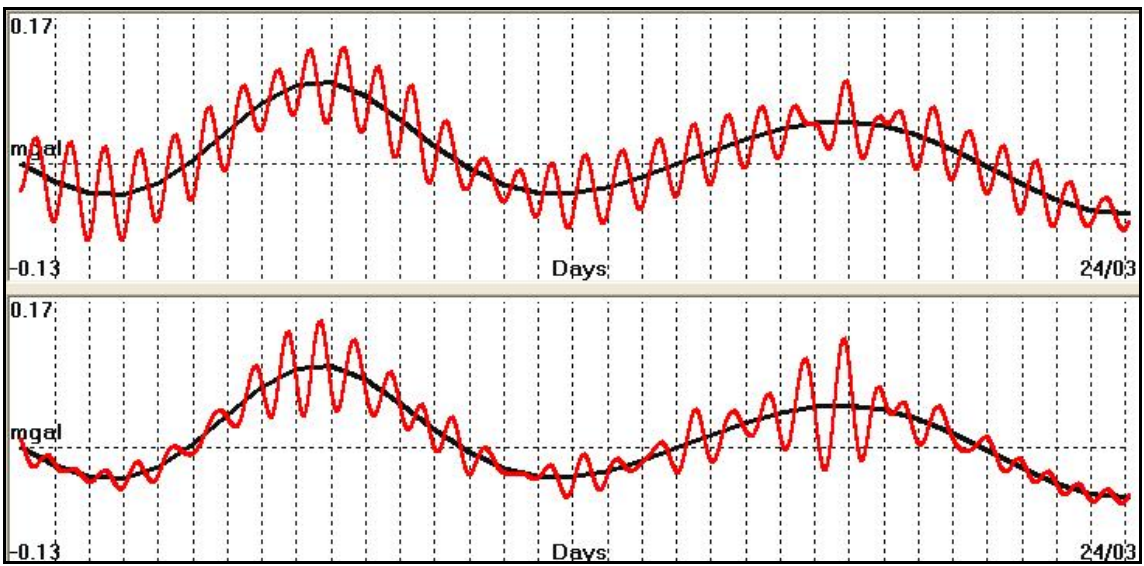

**Fig. 13. Sample of T = 24hours Earth's electric field oscillation vs. M1 (14 days tidal oscillation). Sampling period: 20100220 – 20100324 (YYYYMMDD mode). Upper graph = E – W component, lower graph = N – S component. Two characteristic peaks of the Earth's oscillating electric field are clearly observed on the N – S component which coincide to the M1 (black line) peaks.**

### 2.4. Application on the study period (20080101 to 20100410).

In order to validate the interaction of the piezoelectric mechanism, and the **M1, K1** tidal components so that amplitude modulated Earth's oscillating electric field is generated at the peaks of the **M1** tidal component, a test was run for a rather lengthy period of time of 27 months. The Earth's electric field data recorded by the **ATH** monitoring site in Athens, Greece (Thanassoulas, 2007; www.earthquakeprediction.gr) were used to generate the corresponding oscillating field for the same period of time. To this end an **FFT** band-pass procedure was applied on the raw data. The main target is to test the degree of correlation of the timing of the **M1** tidal peaks to the peak timing of the electric oscillating field. The entire data set has been split in to parts of three months each for easing the presentation. In each graph the green bars indicate the date of the **M1** tidal peak while with red bars represent the Earth's electric oscillating field maxima. Double green and red bars indicate that the corresponding peak did occur in between two days. The results of this correlation are presented in figures (**14 – 22**) as follows:

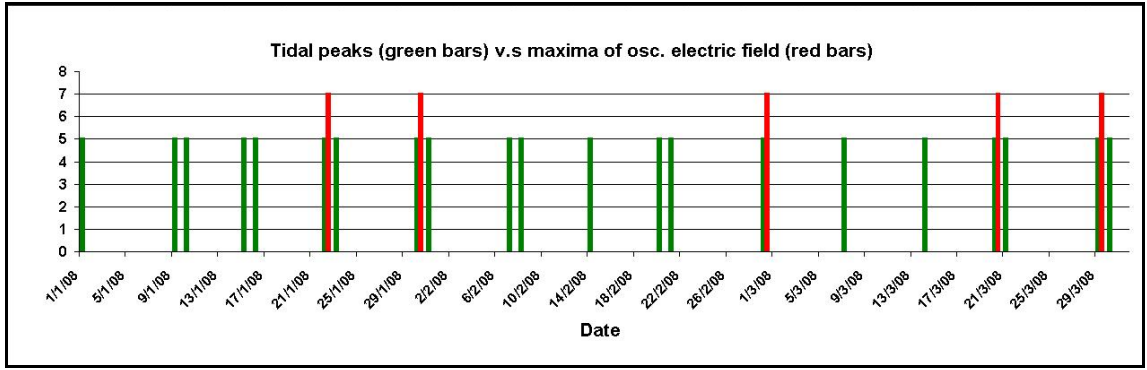

**Fig. 14 . Sampling period 20080101 – 20080331. Green bars represent the M1 tidal peaks.
Red bars represent the maxima of the Earth's oscillating electric field.**

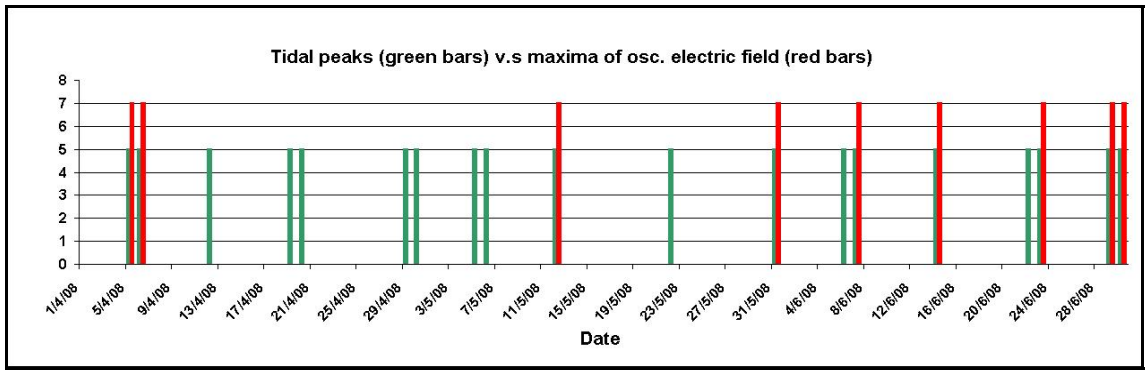

**Fig. 15 . Sampling period 20080401 – 20080630. Green bars represent the M1 tidal peaks.
Red bars represent the maxima of the Earth's oscillating electric field.**

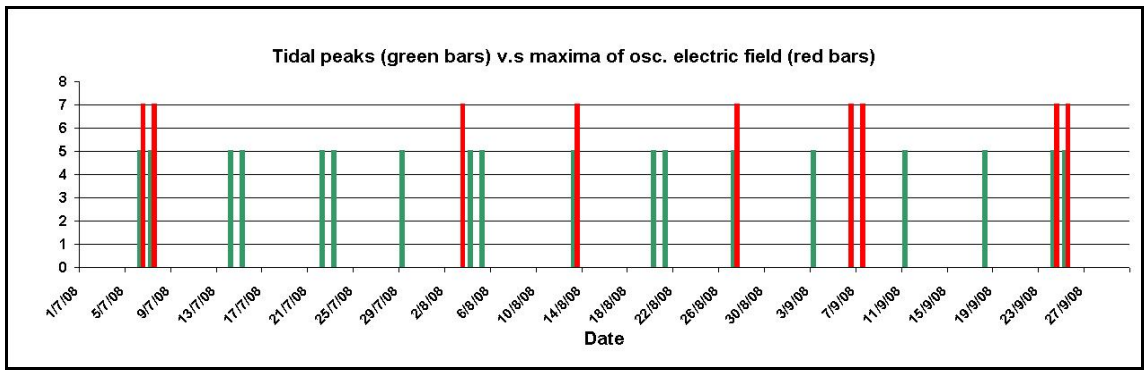

**Fig. 16 . Sampling period 20080701 – 20080930. Green bars represent the M1 tidal peaks.
Red bars represent the maxima of the Earth's oscillating electric field.**

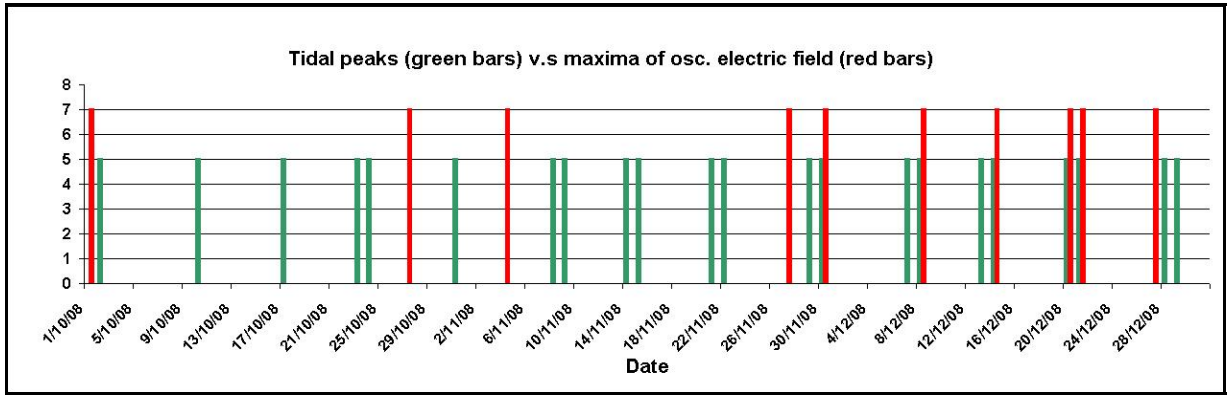

**Fig. 17 . Sampling period 20081001 – 20081231. Green bars represent the M1 tidal peaks.
Red bars represent the maxima of the Earth's oscillating electric field.**

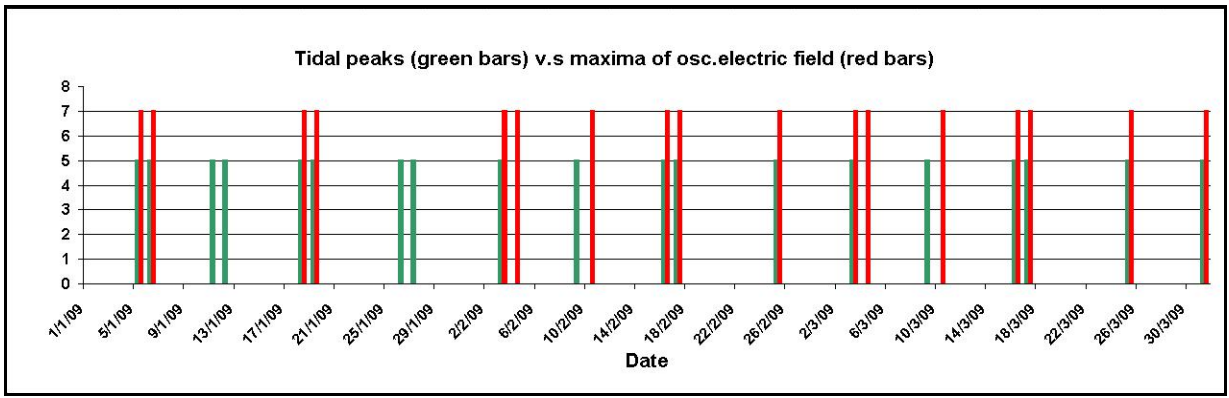

**Fig. 18 . Sampling period 20090101 – 20090331 Green bars represent the M1 tidal peaks.
Red bars represent the maxima of the Earth's oscillating electric field.**

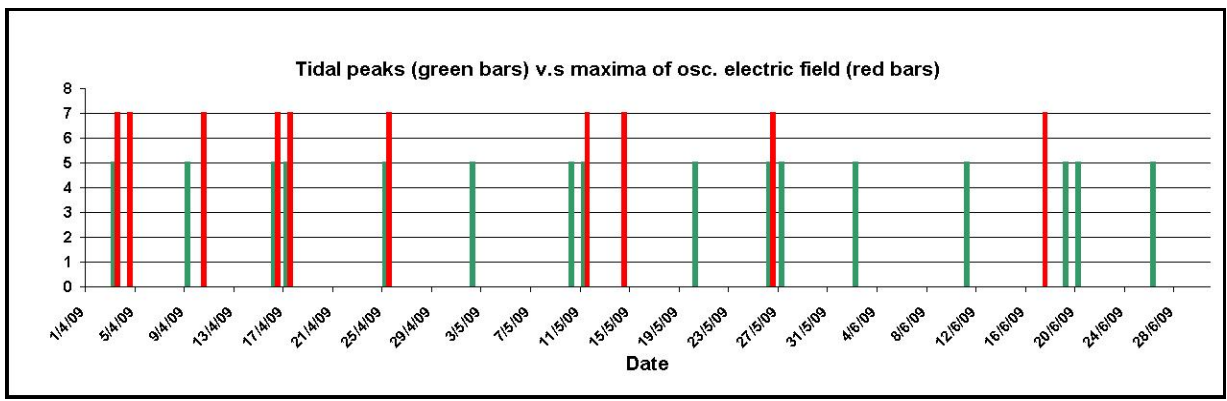

**Fig. 19 . Sampling period 20090401 – 20090630. Green bars represent the M1 tidal peaks.
Red bars represent the maxima of the Earth's oscillating electric field.**

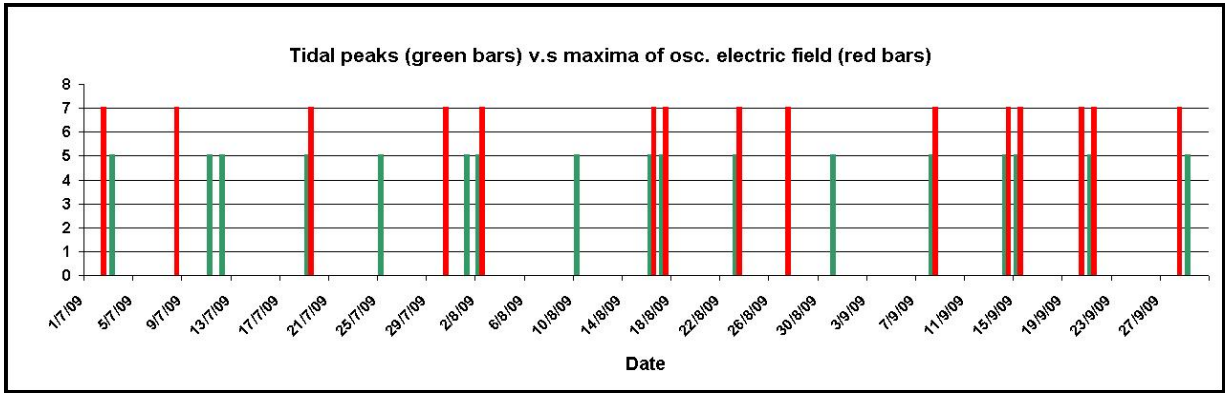

**Fig. 20.** Sampling period 20090701 – 20090930. Green bars represent the **M1** tidal peaks.
Red bars represent the maxima of the Earth's oscillating electric field.

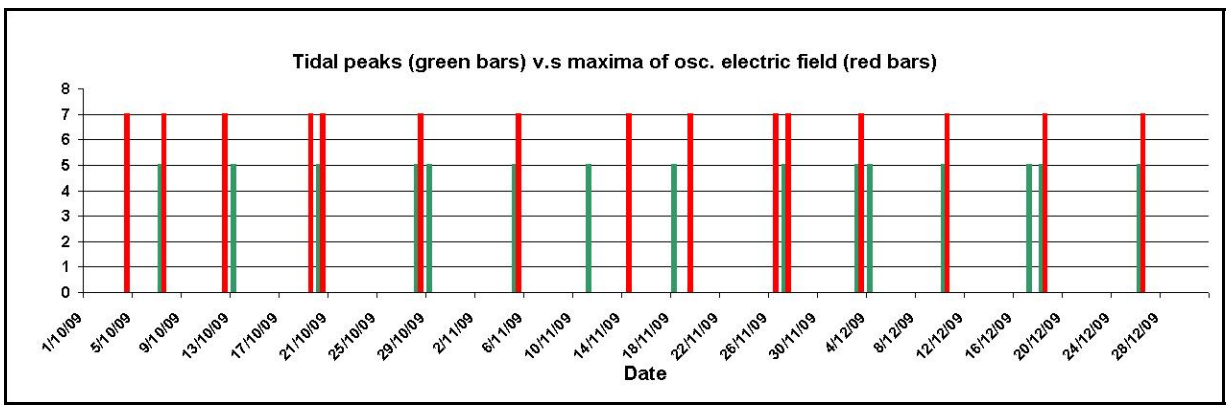

**Fig. 21.** Sampling period 20091001 – 20091231. Green bars represent the **M1** tidal peaks.
Red bars represent the maxima of the Earth's oscillating electric field.

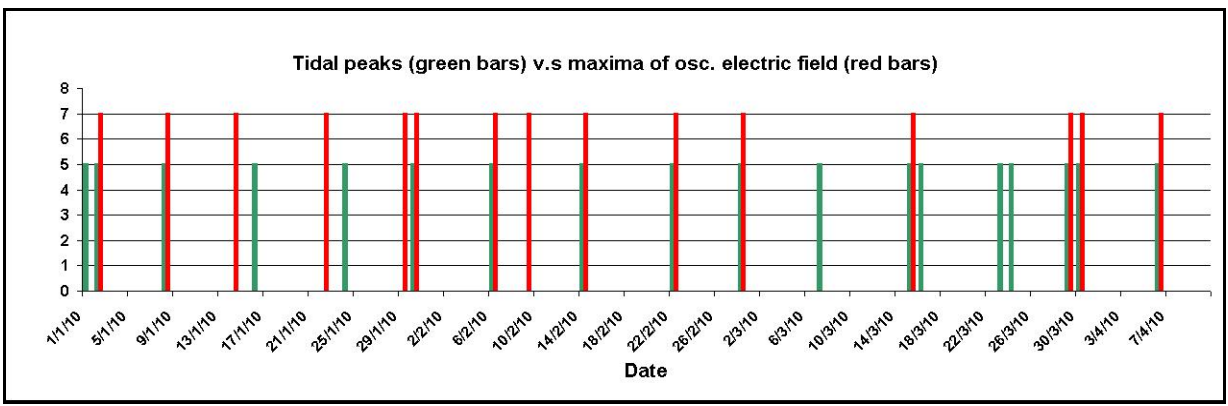

**Fig. 22.** Sampling period 20100101 – 20100410. Green bars represent the **M1** tidal peaks.
Red bars represent the maxima of the Earth's oscillating electric field.

Within the study period 114 **M1** tidal peaks take place. At the same period of time take place 84 peaks of the Earth's oscillating electric field of **T = 24 hours**. Firstly it is assumed that the entire process is totally random. The "by chance" coincidence of an oscillating electric field maximum peak to a **M1** tidal peak is **13.8%** (1/7.2) for a time window of one day (for half wavelength that is two consecutive tidal peaks). The study of the graphs of figures (14) to (22) results in to the following results presented in the graph of figure (23).

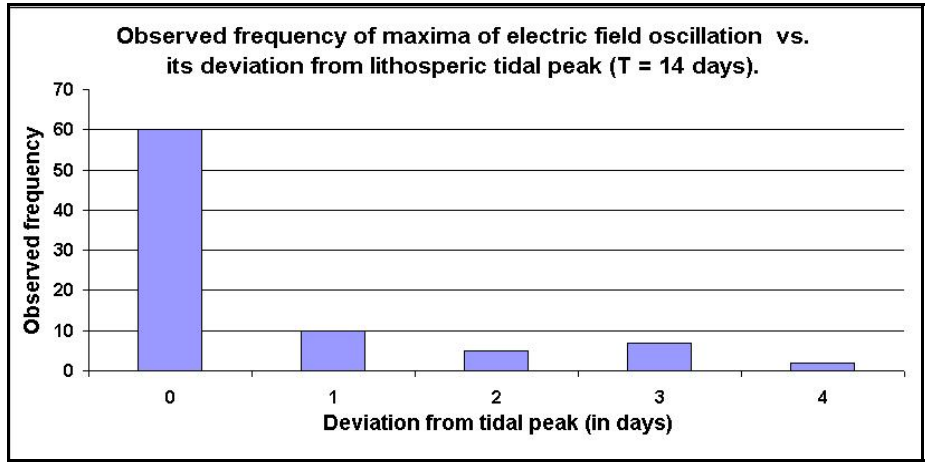

Fig. 23. Observed frequency of maxima of electric field oscillation vs. its deviation from the lithospheric tidal peak oscillation due to **M1** tidal component.

Graph (23) shows that the **71.43% (60/84)** of the observed earth's electric field oscillating maxima were observed exactly on top of the **M1** tidal maxima. That value is by far larger than the one of **13.8%** (by "chance") hit. This value increases to **83.33%** when a time window of +/- 1 day is adopted. These results indicate that the presence of the Earth's oscillating electric field of $T = 1$ day is highly depended on the stress – strain load controlled by the **M1** oscillation.

Furthermore, a comparison is made of the timing of the large EQs **(Ms > 6.0R)** that took place within the study period of time to the timing of the **M1** tidal peaks. The results of this correlation are shown in the following figure (24).

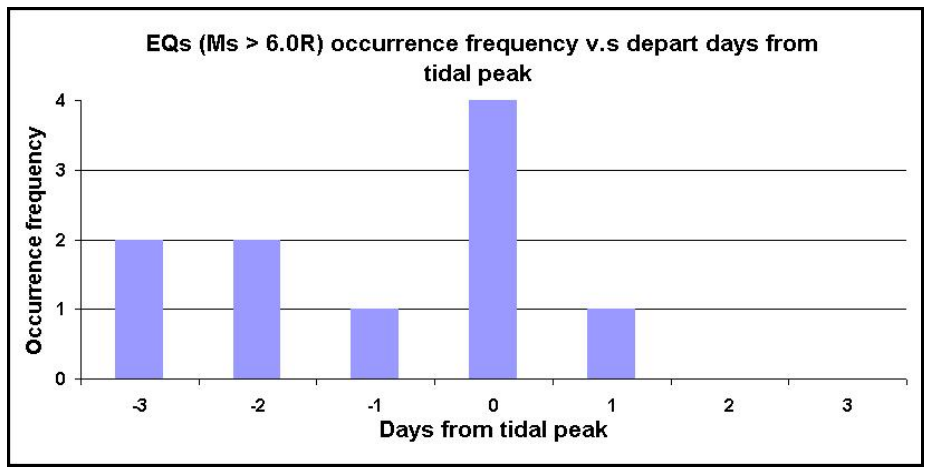

Fig. 24. Comparison of the deviation in days of the occurrence time of large EQs from the corresponding timing of the nearest **M1** tidal peak.

Within the study period, 10 large **(Ms>6R)** took place. Figure (24) shows that, the **40%** of these EQs (in contrast to a **13.8%** for a by "chance" hit) did occur exactly on the peak of the **M1** tidal wave. Although the statistical sample is small it still validates the results of an earlierly presented similar test for a much larger period of some decades of years (Thanassoulas, 2007). If a time window of +/- 1 day is adopted then the coincidence rises up to **60%**.

The results of figure (23) and (24) validate the close relation of the large scale piezoelectric mechanism to the Earth's oscillating electric field and to the triggering of large EQs.

At this time it is worth to point out that the large and small scale piezoelectric mechanism, that is postulated and is activated in the focal region, justifies all of the observed so far preseismic electric signals such as **SES, VLP** (very long period, **oscillating of T = 1/14/365 days**, as well as the Earth's **total electric field** (Thanassoulas 2008a, Thanassoulas et al. 2009).

### 3. Conclusions.

The so called, generally, seismic precursory electric signals have been denied by many researchers. The presented main objection is that in a highly conductive Earth media such signals will dissipate in short distances from the focal area, or what ever has been observed as a precursory signal is quite often of anthropogenic or industrial origin. On the other hand, a quite large number of physical mechanisms have been postulated that are capable of generating precursory electric signals. Details on this topic have been presented by Thanassoulas (2007).

Among the electric signals generating mechanisms is the piezoelectric one. This mechanism was firstly postulated as the generating mechanism of the daily oscillating Earth's electric field component (Thanassoulas et al. 1986, 1993). In the present work a more detailed account for the piezoelectric mechanism is given by taking into consideration the **M1**, in addition to the **K1**, tidal component.

It is shown that the piezoelectric mechanism, triggered at some time before the occurrence of a large EQ, is controlled by both the **M1** and **K1** tidal components. The result of this combined interaction is the generation of a daily oscillating electric field while its amplitude is ordinarily modulated by the **M1** tidal component due to the non-linear strain mechanical – electric properties of the focal area rock formation that take place short before the large EQ occurrence.

The analysis of the data of this work suggested a very good degree of correlation between the timing of the **M1** tidal peaks and the timing of the peak values of the generated by the piezoelectric mechanism oscillating electric field. Therefore, in an indirect way, it consists a warning that a seismogenic area has reached the last critical phase of stress load, when piezoelectricity is triggered, and rock fracturing is expected soon. Moreover, it was shown that the large EQs, which took place in the study period of time, coincide to a large degree to the **M1** tidal peaks.

In general, the piezoelectric mechanism not only justifies the generation of seismic precursory oscillating (of various periods, even of yearly ones; Thanassoulas et al. 2009) electric signals but these signals very often precede the occurrence of large EQs. Moreover, the very same mechanism justifies the generation of **SES**, **VLP** and **Total Earth's electric Field** (complete piezoelectric stress – electric potential curve) often observed before large EQs (Thanassoulas, 2008a).

The same mechanism explains in simple terms of plain physics and rock mechanics when an earthquake will take place and the kind of seismic electric precursory signals that could be generated and observed.

Finally, it seems that the large scale piezoelectric physical model is the most promising one to be used in conjunction to the tidal waves for the large EQ prediction of the occurrence time in short-term mode.

## 4. References.